\documentclass{article}

\evensidemargin=\oddsidemargin
\def\Title#1#2#3{%
    \baselineskip=18pt
    \begin{center}
          {\large\bf{#1} \\ }
          \bigskip\bigskip
          {#2} \\
          {#3} \\
    \end{center}}
\long\def\Abstract#1{%
         \bigskip
         \parbox{0.93\textwidth}{%
                 \begin{center}
                       {\bf Abstract} \\
                 \end{center}
                 \medskip{\baselineskip=14pt #1}
                 \vss}
         \bigskip}

\makeatletter
\renewcommand{\section}%
 {\@startsection{section}{1}{0pt}%
  {-3.25ex plus -1ex minus -.2ex}{1.5ex plus .2ex}%
  {\vspace*{5mm}\raggedright\large\bf }}
\renewcommand{\subsection}%
 {\@startsection{subsection}{2}{0pt}%
  {-2.25ex plus -.5ex minus -.2ex}{-1.5ex plus -.2ex}{\bf }}
\renewcommand{\subsubsection}%
 {\@startsection{subsubsection}{3}{0pt}%
  {-1.25ex plus -.2ex minus -.1ex}{-1.2ex plus -.2ex}{\bf }}
\renewcommand{\thesection}{\arabic{section}.}

\@addtoreset{equation}{section}
\renewcommand{\theequation}{\arabic{section}.\arabic{equation}}
\makeatother

\begin{document}

\Title{Generalized spherically symmetric gravitational model:\\
Hamiltonian dynamics in extended phase space and BRST charge}%
{T. P. Shestakova}%
{Department of Theoretical and Computational Physics,
Southern Federal University,\\
Sorge St. 5, Rostov-on-Don 344090, Russia \\
E-mail: {\tt shestakova@sfedu.ru}}

\Abstract{We construct Hamiltonian dynamics of the generalized spherically symmetric gravitational model in extended phase space. We start from the Faddeev -- Popov effective action with gauge-fixing and ghost terms, making use of gauge conditions in differential form. It enables us to introduce missing velocities into the Lagrangian and then construct a Hamiltonian function according a usual rule which is applied for systems without constraints. The main feature of Hamiltonian dynamics in extended phase space is that it can be proved to be completely equivalent to Lagrangian dynamics derived from the effective action. We find a BRST invariant form of the effective action by adding terms not affecting Lagrangian equations. After all, we construct the BRST charge according to the Noether theorem. Our algorithm differs from that by Batalin, Fradkin and Vilkovisky, but the resulting BRST charge generates correct transformations for all gravitational degrees of freedom including gauge ones. Generalized spherically symmetric model imitates the full gravitational theory much better then models  with finite number of degrees of freedom, so that one can expect appropriate results in the case of the full theory.}

\section{Introduction}
The purpose of this paper is to present Hamiltonian dynamics of the generalized spherically symmetric gravitational model in extended phase space that has two main features: (i) The set of Hamiltonian equations in extended phase space is completely equivalent to the Lagrangian equations obtained from the Faddeev -- Popov effective action for this model. (ii) The BRST charge is derived making use of the Noether theorem and global BRST invariance of the effective action. As a result, the BRST charge generates transformations that coincides with gauge transformations for field variables after change of Grassmannian variables by infinitesimal parameters. These two features distinguish the presented approach from the well-known Batalin -- Fradkin -- Vilkovisky (BFV) approach \cite{BFV1,BFV2,BFV3}. It can be significant when quantizing this model or the full gravitational theory.

As a rule, Hamiltonian formulation for gravity, as a starting point for most attempts to quantize gravity, is constructed following to the Dirac scheme \cite{Dirac1,Dirac2}. Dirac Hamiltonian formulation for gravity \cite{Dirac3} is equivalent to Einstein (Lagrangian) formulation {\it at the level of equations}. It means that Hamiltonian equations for canonical variables (in Dirac's sense) are equivalent to the $\left(i\atop j\right)$ Einstein equations, and the gravitational constraints are equivalent to the $\left(0\atop\mu\right)$ Einstein equations. It is natural to expect that groups of transformations of Lagrangian and Hamiltonian formulations of the same theory are consistent as well, i.e. the group of transformations in Hamiltonian formalism involves gauge transformations of generalized coordinates in the original theory plus transformations of conjugate momenta. However, in Dirac formalism one fails to reproduce transformations for gauge gravitational degrees of freedom $g_{0\mu}$, $\mu=0, 1, 2, 3$. As we shall see below, it is closely connected with the role that given to gauge degrees of freedom, fixing a reference frame, in these two formulations. So, there exist reasons to doubt that Dirac Hamiltonian formulation for gravitational theory can be thought as {\it a fully equivalent} one to the original General Relativity

In Einstein formulation of General Relativity $g_{0\mu}$ components of metric tensor are treated on an equal basis with the rest of components, $g_{ij}$, $i, j=1, 2, 3$, defining 3-space geometry. The theory is invariant under gauge transformations, an infinitesimal form of which is
\begin{equation}
\label{g_transf}
\delta g_{\mu\nu}
 =\int d^4x' R_{\mu\nu |\lambda}(x, x') \eta^{\lambda}(x')
 =\eta^{\lambda}\partial_{\lambda}g_{\mu\nu}
 +g_{\mu\lambda}\partial_{\nu}\eta^{\lambda}
 +g_{\nu\lambda}\partial_{\mu}\eta^{\lambda},
\end{equation}
where $R_{\mu\nu |\lambda}(x, x')$ are generators of gauge transformations and $\eta^{\mu}$ are infinitesimal parameters. The transformations (\ref{g_transf}) are true for all components of metric tensor.  The generators $R_{\mu\nu |\lambda}(x, x')$ satisfy the relations
\begin{equation}
\label{gr.rel}
\int d^4y\left(\frac{\delta R_{\mu\nu |\lambda}(x,x')}{\delta g_{\sigma\tau}(y)}
   R_{\sigma\tau |\rho}(y, x'')
 -\frac{\delta R_{\mu\nu |\rho}(x,x')}{\delta g_{\sigma\tau}(y)}
   R_{\sigma\tau |\lambda}(y,x'')\right)
 =\int d^4y R_{\mu\nu |\sigma}(x,y)C^{\sigma}{}_{\lambda\rho}(y,x',x''),
\end{equation}
$C^{\mu}{}_{\nu\lambda}(x,x',x'')$ are structure functions which are independent on field variables $g_{\mu\nu}$,
\begin{equation}
\label{str.fun}
C^{\mu}{}_{\nu\lambda}(x,x',x'')
 =\delta_{\lambda}^{\mu}\delta(x,x')\partial_{\nu}\delta(x,x'')
 -\delta_{\nu}^{\mu}\delta(x,x'')\partial_{\lambda}\delta(x,x').
\end{equation}
It means that the transformations (\ref{g_transf}) do form a group (see, for example, \cite{DeWitt1}) and the algebra of gauge transformations of General Relativity is closed, according to the terminology of \cite{Hennaux}.

In the Dirac approach only $g_{ij}$ components with their conjugate momenta are included into phase space, while linear combinations of constraints playing the role of generators of transformations in phase space. The so-called momentum constraint produces transformations for $g_{ij}$, which are diffeomorphisms on spacelike 3-surface. But no linear combination of constraints can generate transformations for $g_{0\mu}$ metric components. If linear combinations of constraints only are accepted as generators, one cannot reproduce the full group of transformations (\ref{g_transf}) in Hamiltonian formalism.

An alternative way is to supplement the Dirac approach by some new algorithm of constructing a generator of transformation in phase space. Such an attempt was made, among others, in \cite{Cast}. The Castellani generator proposed in \cite{Cast} indeed produced the transformations for all metric components \cite{KK}. However, we face another problem. Historically, probably because of the complexity of General Relativity, different authors use various parametrizations for gravitational variables. The most known is the ADM parametrization for space-time metric \cite{ADM}:
\begin{equation}
\label{gen.int}
ds^2=\left(-N^2+N_i N^i\right)dt^2+2N_i dt dx^i+g_{ij}dx^i dx^j.
\end{equation}
The ADM parametrization in its reduced form, according to a chosen symmetry, is used in many cosmological models \cite{HH}, black holes models \cite{Kuchar}, etc.

However, in the case of the ADM parametrization the Castellani generator \cite{Cast} produces transformations which will coincide with (\ref{g_transf}) only after the following redefinition of the Castellani infinitesimal parameters $\varepsilon^{\mu}$:
\begin{eqnarray}
\label{redef1}
\eta^0&=&\frac1N\varepsilon^0=\sqrt{(-g^{00})}\varepsilon^0;\\
\label{redef2}
\eta^i&=&\varepsilon^i-\frac{N^i}N\varepsilon^0=\varepsilon^i+\sqrt{(-g^{00})}\frac{g^{0i}}{g^{00}}.
\end{eqnarray}
The necessity for a field-dependent redefinition (\ref{redef1}), (\ref{redef2}) was one of the starting points for the criticism of the ADM formulation in \cite{KK}. One can say that the algorithm \cite{Cast} does not lead to correct transformations in the case of the ADM parametrization. By {\it correct} transformations I mean the ones that follow from (\ref{g_transf}) taking into account a relation between old and new variables and, if some model is considered, its symmetry as well. For example, in the case of the isotropic model with the interval
$ds^2=-N^2(t)dt^2+a^2(t)\left(d\chi^2+\sin^2\chi(d\theta^2+\sin^2\theta d\varphi^2)\right)$
one can easily get from (\ref{g_transf})
\begin{equation}
\label{isotr.tr}
\delta N=-\dot N\eta-N\dot\eta;\qquad
\delta a=-\dot a\eta,\qquad
\eta\equiv\eta^0.
\end{equation}

In \cite{Shest1} it was demonstrated that the Castellani generator fails to produce the transformation (\ref{isotr.tr}) for the lapse function $N$, though it gives a correct transformation (up to a numerical multiplier) for the variable
$\mu=N^2=g_{00}$. In other words, the Castellani algorithm is parametrization-dependent\footnote{In \cite{Shest1} the infinitesimal parameters were denoted as $\theta^{\mu}$ like ghost fields below. It was missed out that $\theta$ denotes $\theta^0$ in the expression for $\delta N$, while in the preceding text in \cite{Shest1} the definition $\theta=\theta_0=g_{00}\theta^0$ was accepted. This mistake could have led to an unfortunate misunderstanding. In \cite{KKK1,FKK} the notion of {\it correct} transformations was criticized, but the criticism itself was based on the wrong expression for $\delta a$.}.

Why does the Castellani algorithm give correct results for the original parametrization in terms of metric components and why does not it for the ADM parametrization? The Castellani procedure has some features that one cannot find in the original formulation of the Dirac approach. Firstly, the generator is constructed as an expansion in terms of derivatives of infinitesimal parameters,
\begin{equation}
\label{gen.gen}
G=\sum\limits_n\varepsilon_{\mu}^{(n)}G_n^{\mu},
\end{equation}
where $G_n^{\mu}$ are first class constraints, $\varepsilon_{\mu}^{(n)}$ are the $n$th order time derivatives of the parameters $\varepsilon_{\mu}$. In principle, it does not contradicts to the Dirac conjecture that generators are linear combinations of constraints, though Dirac never included derivatives of gauge parameters into the linear combinations. Nevertheless, these very derivatives are responsible for the correct result. Secondly, the canonical Hamiltonian used by Dirac \cite{Dirac3} is replaced by the so-called total Hamiltonian which differs for the former by terms $\pi^{0\mu}\dot g_{0\mu}$. Because generalized velocities $\dot g_{0\mu}$ cannot be expressed in terms of metric components and their momenta, the total Hamiltonian depends on the velocities in addition. And, thirdly, since $g_{0\mu}$ and the conjugate momenta $\pi^{0\mu}$ are concerned, the Poisson brackets are determined in {\it formally extended} phase space by including into it these non-canonical variables. It implies significant innovations in comparison with the original Dirac approach, so the Castellani method should be assessed separately from the Dirac procedure.

Poisson brackets algebra and, therefore, constraints' algebra maintain their form under canonical transformation of phase space variables. However, any transformations of gravitational variables, which touch upon gauge degrees of freedom, are not canonical from the viewpoint of the Dirac formalism. In particular, a transition from metric tensor components to the ADM variables should be considered as non-canonical \cite{KK}. In this case the constraints' algebra depends on a parametrization, and it explains why the Castellani procedure, giving correct results for one parametrization, leads to uncorrect results for another. It means that the Castellani algorithm is not general enough to work accurate for any parametrization.

On the other hand, nothing in the Castellani procedure ensures that the infinitesimal parameters $\varepsilon^{\mu}$ involving in (\ref{gen.gen}) must agree with $\eta^{\mu}$ from (\ref{g_transf}) which are infinitesimal coordinate transformations by its origin. The fact that in the Yang -- Mills case the Castellani parameters coincide with gauge group parameters does not mean that the same must take place in any gauge theory. Quite in the spirit of the Dirac approach, Castellani emphasized in \cite{Cast} in his paper the word ``gauge'' was used for transformations which involves arbitrary functions. Since $\varepsilon^{\mu}$ is nothing more than arbitrary functions regardless their role in a gauge theory redefinitions like (\ref{redef1}), (\ref{redef2}) are justified. It explains why the action will be invariant under transformations produced by the Castellani generator though they form a group only in a case when the Castellani parameters are identical with the group ones, otherwise one would need to redefine the parameters. Similar results can be obtained in the framework of the approach \cite{BRR,MS}.

If we understand clearly how the Castellani procedure works, it would be strange to discuss seriously the transformation obtained by this procedure for the ADM formulation, as it was done in \cite{KKK2,KKK3}, to prove that these transformations do not form a group while the diffeomorphism transformations (\ref{g_transf}) do form. In this connection let us mention that the fact that the transformations (\ref{g_transf}) form a group was known very long ago. It was also known that in the original Dirac approach transformations generated by any linear combinations of gravitational constraints do not form a group because structure functions of the constraints' algebra depend on field variables. It is valid for the ADM constraints \cite{DeWitt2}, as well for the Dirac constraints \cite{KK}. It means that the application of the Dirac scheme to gravity gives a theory with an open algebra, in spite of the fact that the algebra of General Relativity is closed. So, we actually deal with {\it two non-equivalent theories with different groups of transformations}. It was emphasized yet in \cite{BFV1} that in the gravitational theory the gauge transformations cannot be presented as canonical transformations, and thus they differ from transformations generated by constraints. In the BFV approach it leads to a new type of additional Feynman diagrams corresponding to four-ghosts interaction which cannot result from the effective action in the Lagrangian form \cite{BFV2}. The new type of diagram does not play an important role while one is interested in only gauge invariant sector in the $S$-matrix theory, for which the BFV approach was originally proposed. However, we can think of it as a considerable mathematical indication that Lagrangian and Hamiltonian formalism appear to be non-equivalent for the full theory of gravity when one deals with spacetime manifolds of any topology, in particular, without asymptotic states which are implied in the $S$-matrix approach.

The central part in the BFV approach is given to the BRST charge constructed as a series in powers of Grassmannian variables with coefficients given by generalized structure functions of constraints algebra \cite{Hennaux}.
\begin{equation}
\label{gen_BFV}
\Omega=c^{\alpha}U^{(0)}_{\alpha}
 +c^{\beta}c^{\gamma}U^{(1)\alpha}_{\gamma\beta}\bar\rho_{\alpha}+\ldots
\end{equation}
$c^{\alpha}$, $\bar\rho_{\alpha}$ are BFV ghosts, $U^{(n)}$ are $n$th order structure functions, while zero order structure functions $U^{(0)}_{\alpha}$ are Dirac secondary constraints. Like the constraints, the BRST charge does not generates correct gauge transformations for all gravitational degrees of freedom including gauge ones. It is not surprising because the form of the BRST charge is determined by constraints algebra and, as was already mentioned, gauge transformations differ from those generated by the constraints.

The purpose of this paper is to present a Hamiltonian dynamics which is free from the shortcomings mentioned above and can be thought of as a real alternative for Dirac generalized Hamiltonian dynamics, as well for the BFV formalism. It has been shown in \cite{Shest2,Shest1} that a transformation of the ADM type is canonical in extended phase space if the extension of phase space implies not just formal including gauge degrees of freedom into it, but also introducing missing velocities by means of gauge conditions in a special (differential) form. It is supposed that the gauge conditions are involved into an effective action, and Hamiltonian dynamics for gravity should be formulated in extended phase space. The proposed approach has already been demonstrated for gravitational models with finite degrees of freedom (\cite{SSV1,SSV2,Shest1} and other papers).

In this work we apply our approach to generalized spherically symmetric gravitational model which imitates the full gravitational theory much better, so that one can see the way how one can get appropriate results in the case of the full theory. In Section 2 the Lagrangian and Hamiltonian dynamics in extended phase space are derived from the effective action for the model, and the structure of the Hamiltonian function and Hamiltonian equations is analyzed. Our main result is presented in Section 3 where we shall make use of BRST invariance of effective action and construct the BRST charge according to the Noether theorem which generates correct transformations for all gravitational degrees of freedom. The proposed method will be shown to be self-consistent, and the equivalence of the Lagrangian and Hamiltonian formulations can be proved by direct calculations.

\section{The model, its Lagrangian and Hamiltonian dynamics}
In this paper we shall follow to the ADM parametrization (\ref{gen.int}). Under the condition of spherical symmetry the metric is reduced to
\begin{eqnarray}
ds^2&=&\left[-N^2(t,r)+(N^r(t,r))^2V^2(t,r)\right]dt^2+2N^r(t,r)V^2(t,r)dt dr\nonumber\\
\label{mod.int}
&+&V^2(t,r)dr^2+W^2(t,r)\left(d\theta^2+\sin^2\theta d\varphi^2\right).
\end{eqnarray}
where $N^r=N^1$ is the only component of the shift vector. In this model we have two gauge variables $N$ and $N^r$ which are fixed by two gauge conditions
\begin{eqnarray}
\label{g.c.1}
N&=&f(V,W);\\
\label{g.c.2}
N^r&=&f^r(V,W).
\end{eqnarray}
$f(V,W)$, $f^r(V,W)$ are arbitrary functions. In differential form the gauge conditions will introduce missing velocities into the effective Lagrangian, so ensuring {\it an actual extension of phase space}:
\begin{eqnarray}
\label{d.c.1}
\dot N&=&\frac{\partial f}{\partial V}\dot V+\frac{\partial f}{\partial W}\dot W;\\
\label{d.c.2}
\dot N^r&=&\frac{\partial f^r}{\partial V}\dot V+\frac{\partial f^r}{\partial W}\dot W.
\end{eqnarray}

We shall consider the Faddeev -- Popov effective action including gauge and ghost sectors as it appears in the path integral approach to gauge field theories \cite{FP},
\begin{equation}
\label{full-act1}
S_{(eff)}=S_{(grav)}+S_{(gauge)}+S_{(ghost)}.
\end{equation}

The gravitational part of the effective action
\begin{equation}
\label{grav-act1}
S_{(grav)}=\int d^4x\sqrt{-g}R
\end{equation}
is invariant under gauge transformations (\ref{g_transf}), but the gravitational Lagrangian involves second derivatives of metric components. To get field equations it is much easier to make use of the Lagrangian which is quadratic in first derivatives of metric components and can be obtained from the original one by omitting total derivatives. However, we shall have to return to the original gravitational Lagrangian when deriving the BRST charge in accordance with the Noether theorem (see Section 3).

The gauge-fixing part of the action is
\begin{equation}
\label{gfix-act1}
S_{(gauge)}
 =\int dt\int\limits_0^{\infty}dr\left[\lambda_0\left(\dot N-\frac{\partial f}{\partial V}\dot V
   -\frac{\partial f}{\partial W}\dot W\right)
  +\lambda_r\left(\dot N^r-\frac{\partial f^r}{\partial V}\dot V
   -\frac{\partial f^r}{\partial W}\dot W\right)\right].
\end{equation}

Taking into account gauge transformations for gravitational variables
\begin{eqnarray}
\label{delN}
\delta N&=&-\dot N\eta^0-N'\eta^r-N\dot\eta^0+N N^r(\eta^0)';\\
\label{delNr}
\delta N^r&=&-\dot N^r\eta^0-(N^r)'\eta^r-N^r\dot\eta^0-\dot\eta^r
   +N^r(\eta^r)'+\frac{N^2}{V^2}(\eta^0)'+(N^r)^2(\eta^0)';\\
\label{delV}
\delta V&=&-\dot V\eta^0-V'\eta^r-V(\eta^r)'-V N^r(\eta^0)';\\
\label{delW}
\delta W&=&-\dot W\eta^0-W'\eta^r,
\end{eqnarray}
that follow from (\ref{g_transf}), we get the Faddeev -- Popov ghost action in the form
\begin{eqnarray}
\label{ghost-act1}
S_{(ghost)}&=&\int dt\int\limits_0^{\infty}dr
   \left[\bar\theta_0\frac{d}{dt}\left(-\dot N\theta^0-N'\theta^r-N\dot\theta^0
    +N N^r(\theta^0)'\right.\right.\nonumber\\
&-&\left.\frac{\partial f}{\partial V}\left[-\dot V\theta^0-V'\theta^r-V(\theta^r)'-V N^r(\theta^0)'\right]
    -\frac{\partial f}{\partial W}\left[-\dot W\theta^0-W'\theta^r\right]\right)\nonumber\\
&+&\bar\theta_r\frac{d}{dt}\left(-\dot N^r\theta^0-(N^r)'\theta^r-N^r\dot\theta^0-\dot\theta^r
    +N^r(\theta^r)'+\frac{N^2}{V^2}(\theta^0)'+(N^r)^2(\theta^0)'\right.\nonumber\\
&-&\left.\left.\frac{\partial f^r}{\partial V}\left[-\dot V\theta^0-V'\theta^r-V(\theta^r)'-V N^r(\theta^0)'\right]
    -\frac{\partial f^r}{\partial W}\left[-\dot W\theta^0-W'\theta^r\right]\right)\right]
\end{eqnarray}
$\bar\theta_0$, $\theta^0$, $\bar\theta_r$, $\theta^r$ are ghost variables. After redefinition
\begin{equation}
\label{lambda}
\pi_N=\lambda_0+\dot{\bar\theta_0}\theta^0;\quad
\pi_{N^r}=\lambda_r+\dot{\bar\theta_r}\theta^0
\end{equation}
we can write the effective Lagrangian in the form without second derivatives:
\begin{eqnarray}
\label{full-act2}
S_{(eff)}&=&\int dt\int\limits_0^{\infty}dr\left(\frac{\dot V\dot W W}N
  +\frac{V\dot W^2}{2N}
  -\frac{N'W'W}V
  -\frac{N(W')^2}{2V}
  -\frac{NV}2\right.\nonumber\\
 &-&\frac{W'\dot WVN^r}N
  -\frac{W W'\dot V N^r}N
  -\frac{W\dot W V'N^r}N
  -\frac{W\dot W V(N^r)'}N\nonumber\\
 &+&\frac{W W'V'(N^r)^2}N
  + \frac{W W'V N^r(N^r)'}N
  +\frac{(W')^2 V(N^r)^2}{2N}\nonumber\\
 &+&\pi_N\left(\dot N-\frac{\partial f}{\partial V}\dot V-\frac{\partial f}{\partial W}\dot W\right)
  +\pi_{N^r}\left(\dot N^r-\frac{\partial f^r}{\partial V}\dot V
   -\frac{\partial f^r}{\partial W}\dot W\right)\nonumber\\
 &+&\dot{\bar\theta_0}\theta^r\left(N'-\frac{\partial f}{\partial V}V'
   -\frac{\partial f}{\partial W}W'\right)\nonumber\\
 &+&\dot{\bar\theta_0}\left(N\dot\theta^0-NN^r(\theta^0)'-\frac{\partial f}{\partial V}VN^r(\theta^0)'
   -\frac{\partial f}{\partial V}V(\theta^r)'\right)\nonumber\\
 &+&\dot{\bar\theta_r}\left[N^r\dot\theta^0-\left(\frac{N^2}{V^2}+(N^r)^2\right)(\theta^0)'+\dot\theta^r
   -N^r(\theta^r)'+(N^r)'\theta^r\right.\nonumber\\
 &-&\left.\left.\frac{\partial f^r}{\partial V}\left(VN^r(\theta^0)'+V(\theta^r)'+V'\theta^r\right)
   -\frac{\partial f^r}{\partial W}W'\theta^r\right]\right)
\end{eqnarray}

Variation of the effective action with respect to $N$, $N^r$, $V$, $W$ yields the Einstein equations for the model with additional terms resulting from the gauge-fixing and ghost parts of the action. These equations can be called {\it the gauged Einstein equations}. By adding ghost equations and gauge conditions (\ref{d.c.1}), (\ref{d.c.2}) to the gauged Einstein equations, we obtain {\it the extended set of Lagrangian equations} for our model which is presented in Appendix A.

Now we can find the momenta conjugate to all gravitational and ghost variables:
\begin{eqnarray}
\label{PN}
P_N&=&\pi_N;\\
\label{PNr}
P_{N^r}&=&\pi_{N^r};\\
\label{PV}
P_V&=&\frac{W\dot W}N-\frac{W'W N^r}N-\pi_N\frac{\partial f}{\partial V}
   -\pi_{N^r}\frac{\partial f^r}{\partial V};\\
\label{PW}
P_W&=&\frac{W\dot V}N+\frac{V\dot W}N-\frac{W'V N^r}N-\frac{W V'N^r}N-\frac{V W(N^r)'}N
   -\pi_N\frac{\partial f}{\partial W}-\pi_{N^r}\frac{\partial f^r}{\partial W};\\
\label{Pbt0}
P_{\bar\theta_0}&=&N'\theta^r-\frac{\partial f}{\partial V}V'\theta^r
   -\frac{\partial f}{\partial W}W'\theta^r
   +N\dot\theta^0-N N^r(\theta^0)'-\frac{\partial f}{\partial V}V N^r(\theta^0)'
   -\frac{\partial f}{\partial V}V(\theta^r)';\\
\label{Pt0}
\bar P_{\theta^0}&=&\dot{\bar\theta_0}N+\dot{\bar\theta_r}N^r;\\
P_{\bar\theta_r}&=&N^r\dot\theta^0-\frac{N^2}{V^2}(\theta^0)'-(N^r)^2(\theta^0)'
   +\dot\theta^r-N^r(\theta^r)'+(N^r)'\theta^r\nonumber\\
\label{Pbtr}
&-&\frac{\partial f^r}{\partial V}V N^r(\theta^0)'
   -\frac{\partial f^r}{\partial V}V(\theta^r)'
   -\frac{\partial f^r}{\partial V}V'\theta^r
   -\frac{\partial f^r}{\partial W}W'\theta^r;\\
\label{Ptr}
\bar P_{\theta^r}&=&\dot{\bar\theta_r}.
\end{eqnarray}

Introducing of the missing velocities by means of the differential form of gauge conditions (\ref{d.c.1}), (\ref{d.c.2}) enables us to construct a Hamiltonian in extended phase space not applying to the Dirac procedure, by the usual rule
\begin{equation}
\label{Ham1}
H=\int\limits_0^{\infty}dr\left(P_N\dot N+P_{N^r}\dot N^r+P_V\dot V+P_W\dot W
   +\bar P_{\theta^0}\dot\theta^0+\dot{\bar\theta_0}P_{\bar\theta_0}
   +\bar P_{\theta^r}\dot\theta^r+\dot{\bar\theta_r}P_{\bar\theta_r}-L\right);
\end{equation}
\begin{eqnarray}
H&=&\int\limits_0^{\infty}dr\left[\frac N W P_V P_W
   -\frac{N V}{2W^2}P_V^2+P_V V'N^r+P_V V(N^r)'+P_W W'N^r
   +\frac{N'W'W}V+\frac{N(W')^2}{2V}+\frac{N V}2\right.\nonumber\\
&+&P_N\frac{\partial f}{\partial V}V'N^r
   +P_N\frac{\partial f}{\partial W}W'N^r
   +P_N\frac{\partial f}{\partial V}V(N^r)'
   +P_{N^r}\frac{\partial f^r}{\partial V}V'N^r
   +P_{N^r}\frac{\partial f^r}{\partial W}W'N^r\nonumber\\
&+&P_{N^r}\frac{\partial f^r}{\partial V}V(N^r)'
   +\frac N W P_V P_N\frac{\partial f}{\partial W}
   +\frac N W P_W P_N\frac{\partial f}{\partial V}
   -\frac{N V}{W^2}P_V P_N\frac{\partial f}{\partial V}\nonumber\\
&+&\frac N W P_V P_{N^r}\frac{\partial f^r}{\partial W}
   +\frac N W P_W P_{N^r}\frac{\partial f^r}{\partial V}
   -\frac{N V}{W^2}P_V P_{N^r}\frac{\partial f^r}{\partial V}
   -\frac{N V}{2W^2} P_N^2\left(\frac{\partial f}{\partial V}\right)^2\nonumber\\
&-&\frac{N V}{2W^2}P_{N^r}^2\left(\frac{\partial f^r}{\partial V}\right)^2
   +\frac N W P_N^2\frac{\partial f}{\partial V}
    \frac{\partial f}{\partial W}
   +\frac N W P_{N^r}^2\frac{\partial f^r}{\partial V}
    \frac{\partial f^r}{\partial W}\nonumber\\
&+&\frac N W P_N P_{N^r}\frac{\partial f}{\partial V}
    \frac{\partial f^r}{\partial W}
   +\frac N W P_N P_{N^r}\frac{\partial f}{\partial W}
    \frac{\partial f^r}{\partial V}
   -\frac{N V}{W^2}P_N P_{N^r}\frac{\partial f}{\partial V}
    \frac{\partial f^r}{\partial V}\nonumber\\
&+&\frac1N\bar P_{\theta^0}P_{\bar\theta_0}
   +\bar P_{\theta^r}P_{\bar\theta_r}
   -\frac{N^r}N\bar P_{\theta^r}P_{\bar\theta_0}
   -\frac{N'}N\bar P_{\theta^0}\theta^r
   +N^r\bar P_{\theta^0}(\theta^0)'
   +N^r\bar P_{\theta^r}(\theta^r)'\nonumber\\
&-&(N^r)'\bar P_{\theta^r}\theta^r
   +\frac{N'N^r}N\bar P_{\theta^r}\theta^r
   +\frac{N^2}{V^2}\bar P_{\theta^r}(\theta^0)'
   +\frac{\partial f^r}{\partial V}V'\bar P_{\theta^r}\theta^r
   +\frac{\partial f^r}{\partial W}W'\bar P_{\theta^r}\theta^r\nonumber\\
&+&\frac{\partial f^r}{\partial V}V N^r\bar P_{\theta^r}(\theta^0)'
   +\frac{\partial f^r}{\partial V}V\bar P_{\theta^r}(\theta^r)'
   +\frac V N\frac{\partial f}{\partial V}\bar P_{\theta^0}(\theta^r)'
   +\frac{V N^r}N\frac{\partial f}{\partial V}\bar P_{\theta^0}(\theta^0)'\nonumber\\
&-&\frac{V N^r}N\frac{\partial f}{\partial V}\bar P_{\theta^r}(\theta^r)'
   -\frac{V(N^r)^2}N\frac{\partial f}{\partial V}\bar P_{\theta^r}(\theta^0)'
   +\frac{V'}N\frac{\partial f}{\partial V}\bar P_{\theta^0}\theta^r\nonumber\\
\label{Ham-EPS}
&-&\left.\frac{V'N^r}N\frac{\partial f}{\partial V}\bar P_{\theta^r}\theta^r
   +\frac{W'}N\frac{\partial f}{\partial W}\bar P_{\theta^0}\theta^r
   -\frac{W'N^r}N\frac{\partial f}{\partial W}\bar P_{\theta^r}\theta^r\right].
\end{eqnarray}

The first line in (\ref{Ham-EPS}) is the Hamiltonian for pure gravity that can be presented as a linear combination of Dirac secondary constraints since it is believed that a full derivative with respect to $r$ can be omitted in this expression:
\begin{equation}
\label{Ham-D}
H_D=\int\limits_0^{\infty}dr\left[N\left(\frac1W P_V P_W-\frac V{2W^2}P_V^2
   -\frac{W W''}V-\frac{(W')^2}{2V}+\frac{V'W W'}{V^2}+\frac V2\right)
   +N^r\left(P_W W'-P'_V V\right)\right].
\end{equation}
However, as it follows from (\ref{Ham-EPS}), the Hamiltonian in extended phase space cannot be written down as a linear combination of constraints. Now we can write down the set of Hamiltonian equations in extended phase space presented explicitly in Appendix B.

It is important to emphasized that in this formulation of Hamiltonian dynamics the constraints as well as the gauge conditions have the status of Hamiltonian equations. Indeed, the Hamiltonian equations (\ref{DN}), (\ref{DNr}) coincide with the gauge conditions (\ref{d.c.11}), (\ref{d.c.22}), while the equations (\ref{DPN}), (\ref{DPNr}) reproduce the constraints (\ref{eLN}), (\ref{eLNr}) in the Lagrangian formalism. The equations (\ref{DV}) -- (\ref{DPW}) for physical gravitational degrees of freedom after some rearrangement can be shown to be equivalent to the dynamical Lagrangian equations (\ref{eLV}), (\ref{eLW}), and Eqs. (\ref{Dt0}) -- (\ref{DbPtr}) are equivalent to the ghost equations (\ref{eLbt0}) -- (\ref{eLtr}). Thus, in this Section we have got two sets of extended equations for our spherically symmetric model in the Lagrangian and Hamiltonian formalisms, which are proved to be completely equivalent.

\section{The BRST charge}
It is known that the Faddeev -- Popov effective action possesses a residual global symmetry, the so-called BRST symmetry. In the Lagrangian formalism the BRST transformations for our model are given by (\ref{delN}) -- (\ref{delW}), where infinitesimal parameters $\eta^{\mu}$ should be replaced by $\bar\varepsilon\theta^{\mu}$, $\bar\varepsilon$ is a Grassmannian parameter,
\begin{eqnarray}
\label{delN1}
\delta N&=&\bar\varepsilon\left[-\dot N\theta^0-N'\theta^r-N\dot\theta^0+N N^r(\theta^0)'\right];\\
\label{delNr1}
\delta N^r&=&\bar\varepsilon\left[-\dot N^r\theta^0-(N^r)'\theta^r-N^r\dot\theta^0-\dot\theta^r
   +N^r(\theta^r)'+\frac{N^2}{V^2}(\theta^0)'+(N^r)^2(\theta^0)'\right];\\
\label{delV1}
\delta V&=&\bar\varepsilon\left[-\dot V\theta^0-V'\theta^r-V(\theta^r)'-V N^r(\theta^0)'\right];\\
\label{delW1}
\delta W&=&\bar\varepsilon\left[-\dot W\theta^0-W'\theta^r\right].
\end{eqnarray}
Moreover,
\begin{eqnarray}
\label{delt0}
\delta\theta^0&=&\bar\varepsilon\left[\dot\theta^0\theta^0+(\theta^0)'\theta^r\right];\\
\label{deltr}
\delta\theta^r&=&\bar\varepsilon\left[\dot\theta^r\theta^0+(\theta^r)'\theta^r\right];\\
\label{delbt0}
\delta\bar\theta_0&=&-\bar\varepsilon\lambda_0;\\
\label{delbtr}
\delta\bar\theta_r&=&-\bar\varepsilon\lambda_r;\\
\label{dell0}
\delta\lambda_0&=&0;\\
\label{dellr}
\delta\lambda_r&=&0.
\end{eqnarray}
The transformations (\ref{delN1}) -- (\ref{dellr}) should be supplemented by coordinated transformations
\begin{equation}
\label{coord.tr}
\delta t=\bar\varepsilon\theta^0;\qquad
\delta r=\bar\varepsilon\theta^r.
\end{equation}
As a consequence of a global symmetry there exists a BRST charge which plays a role of a generator of BRST transformations in extended phase space. As we have already mentioned, in the BFV approach it is constructed as a series in powers of Grassmannian variables with coefficients given by generalized structure functions of constraints algebra (\ref{gen_BFV}). Since the BFV prescription of constructing the BRST charge is essentially rely upon constraints algebra, it cannot produce correct transformations for gauge gravitational variables, like a linear combination of constraints cannot produce them in the Dirac approach. In \cite{Shest1} we have analyzed the Castellani algorithm \cite{Cast} that aims at modifying the Dirac scheme and constructing a generator producing correct transformations for all variables. However, this algorithm fails to be applied to an arbitrary parametrization of gravitational variables, so it is not general enough and cannot be considered as a required solution to the problem.

At the same time, the existence of global BRST symmetry enables us to propose a method based upon the Noether theorem and the equivalence of Lagrangian dynamics and Hamiltonian dynamics in extended phase space. Until now it was applied to models with finite numbers degrees of freedom \cite{SSV2,Shest1}. The method is straightforward, although it requires rather tedious calculations for more complicated models. Recently in \cite{CM} another approach to construct a BRST charge has been put forward. This approach implies some modification of the Noether procedure. The authors applied it to the Friedmann -- Robertson -- Walker model and reproduced the result of \cite{Shest1}. It is not clear if the proposed modification of the Noether procedure could give a significant simplification of calculations.

In this section we shall apply our straightforward method to the spherically symmetric model. Let us note that nothing prevent one from applying it to any other gravitational model including the full theory of gravity. The fact that gauge degrees of freedom are treated on the equal footing with other variables allows one to make transformations of variables including gauge ones which have been proved to be canonical in extended phase space \cite{Shest1} and do not affect the algebra of Poisson brackets. So, the proposed method will work for any reasonable parametrization of gravitational variables.

You can find the proof of BRST symmetry of the Faddeev -- Popov effective action for Yang -- Mills fields in any book on quantum field theory \cite{Weinberg}. In the case of gravity we deal with space-time symmetry, and we should take into account explicit dependence of the Lagrangian and the measure on space-time coordinates. In the early papers on Quantum Gravity enough attention was not been paid to this circumstance which singles out gravity among other gauge fields (see \cite{DeWitt1} and references therein). One can check that the sum of gauge-fixing and ghost parts of the action (\ref{gfix-act1}), (\ref{ghost-act1}) is not invariant under transformations (\ref{delN1}) -- (\ref{coord.tr}). In some works the BRST invariance is guaranteed by asymptotical boundary conditions for ghosts and Lagrange multipliers \cite{Hennaux,Hall}. The legitimacy of asymptotic boundary conditions is questionable in the case of space-time of arbitrary topology. Therefore, we seek for a BRST invariant form of the action without appealing to any additional conditions. One can check that to ensure its BRST invariance we have to add to the action the following term containing only full derivatives and not affecting the set of equations obtained in Section 2:
\begin{eqnarray}
S_{(add)}&=&\int dt\int\limits_0^{\infty}dr
   \left(\frac d{dt}\left[\bar\theta_0\left(\dot N-\frac{\partial f}{\partial V}\dot V
     -\frac{\partial f}{\partial W}\dot W\right)\theta^0\right]
    +\frac d{dr}\left[\bar\theta_0\left(\dot N-\frac{\partial f}{\partial V}\dot V
     -\frac{\partial f}{\partial W}\dot W\right)\theta^r\right]\right.\nonumber\\
\label{add-act}
&+&\left.\frac d{dt}\left[\bar\theta_r\left(\dot N^r-\frac{\partial f^r}{\partial V}\dot V
     -\frac{\partial f^r}{\partial W}\dot W\right)\theta^0\right]
    +\frac d{dr}\left[\bar\theta_r\left(\dot N^r-\frac{\partial f^r}{\partial V}\dot V
     -\frac{\partial f^r}{\partial W}\dot W\right)\theta^r\right]\right).
\end{eqnarray}

As was mentioned in Section 2, the gravitational part of the action in (\ref{full-act2}) is not invariant under gauge transformation and, therefore under BRST transformations (\ref{delN1}) -- (\ref{delW1}). Then, we should return to the gravitational action (\ref{grav-act1}). Now we deal with the Lagrangian which involves second derivatives of metric components and ghosts. The BRST charge is constructed in accordance with the Noether theorem generalized for theories with high derivatives:
\begin{equation}
\label{Noet.BRST}
\Omega=\int d^3x\left[\frac{\partial L}{\partial(\partial_0\phi^a)}\delta\phi^a
   +\frac{\partial L}{\partial(\partial _0\partial_{\mu}\phi^a)}\delta(\partial_{\mu}\phi^a)
   -\partial_{\mu}\left(\frac{\partial L}{\partial(\partial_0\partial_{\mu}\phi^a)}\right)\delta\phi^a
   +\partial _0\left(L x^0\right)\right].
\end{equation}
$\phi^a$ stands for all variables $N$, $N^r$, $V$, $W$ and ghosts. After some tedious calculations we come to the following expression for the BRST charge in the spherically symmetric model:
\begin{eqnarray}
\Omega&=&\int\!dr\left[-{\cal H}\theta^0-P_V V'\theta^r
   -P_N\frac{\partial f}{\partial V}V'\theta^r
   -P_{N^r}\frac{\partial f^r}{\partial V}V'\theta^r\right.\nonumber\\
&-&P_W W'\theta^r-P_N\frac{\partial f}{\partial W}W'\theta^r
   -P_{N^r}\frac{\partial f^r}{\partial W}W'\theta^r\nonumber\\
&-&P_V V N^r(\theta^0)'-P_N\frac{\partial f}{\partial V} V N^r(\theta^0)'
   -P_{N^r}\frac{\partial f^r}{\partial V} V N^r(\theta^0)'\nonumber\\
&-&P_V V(\theta^r)'-P_N\frac{\partial f}{\partial V} V(\theta^r)'
   -P_{N^r}\frac{\partial f^r}{\partial V} V(\theta^r)'\nonumber\\
\label{BRST1}
&-&\left.\bar P_{\theta^0}(\theta^0)'\theta^r
   -\bar P_{\theta^r}(\theta^r)'\theta^r
   -P_N P_{\bar\theta_0}-P_{N^r}P_{\bar\theta_r}
   -\frac{N W W'(\theta^0)'}V\right],
\end{eqnarray}
$\cal H$ is a Hamiltonian density in (\ref{Ham-EPS}). It can be directly checked that the charge (\ref{BRST1}) generates transformations (\ref{delN1}) -- (\ref{delbtr}). Let us emphasized that the Hamiltonian equations in extended phase space, in particular, constraints and gauge conditions which have the status of Hamiltonian equations, can be used to get correct results, for instance,
\begin{eqnarray}
\delta N&=&\left\{N,\;\bar\varepsilon\Omega\right\}
   =\bar\varepsilon\frac{\delta\Omega}{\delta P_N}\nonumber\\
&=&\bar\varepsilon\left[-\frac{\partial{\cal H}}{\partial P_N}\theta^0
   -\frac{\partial f}{\partial V}V'\theta^r
   -\frac{\partial f}{\partial W}W'\theta^r
   -\frac{\partial f}{\partial V} V N^r(\theta^0)'
   -\frac{\partial f}{\partial V} V(\theta^r)'-P_{\bar\theta_0}\right]\nonumber\\
\label{delN2}
&=&\bar\varepsilon\left[-\dot N\theta^0-N'\theta^r-N\dot\theta^0+N N^r(\theta^0)'\right].
\end{eqnarray}
Here we used one of the Hamiltonian equations
$\dot N=\displaystyle\frac{\delta H}{\delta P_N}$ (\ref{DN}), and the expression for $P_{\bar\theta_0}$ (\ref{Pbt0}). To check (\ref{dell0}), (\ref{dellr}) one should firstly find $\delta P_N$, $\delta P_{N^r}$.
\begin{eqnarray}
\delta P_N&=&\left\{P_N,\;\bar\varepsilon\Omega\right\}
   =-\bar\varepsilon\frac{\delta\Omega}{\delta N}
   =\bar\varepsilon\left[-\frac{\partial\Omega}{\partial N}
    +\left(\frac{\partial\Omega}{\partial N'}\right)'\right]\nonumber\\
&=&\bar\varepsilon\left[\frac{\partial{\cal H}}{\partial N}\theta^0
    -\left(\frac{\partial{\cal H}}{\partial N'}\theta^0\right)'
    +\frac{W W'(\theta^0)'}V\right]\nonumber\\
&=&\bar\varepsilon\left[\left(\frac{\partial{\cal H}}{\partial N}
    -\left(\frac{\partial{\cal H}}{\partial N'}\right)'\right)\theta^0
    -\frac{\partial{\cal H}}{\partial N'}(\theta^0)'
    +\frac{W W'(\theta^0)'}V\right]\nonumber\\
&=&\bar\varepsilon\left[-\dot P_N\theta^0-\left(\frac{W W'}V
    -\frac1N\bar P_{\theta^0}\theta^r
    +\frac{N^r}N\bar P_{\theta^r}\theta^r\right)(\theta^0)'
    +\frac{W W'(\theta^0)'}V\right]\nonumber\\
\label{delPN}
&=&\bar\varepsilon\left[-\dot P_N\theta^0
    +\frac1N\left(N\dot{\bar\theta_0}+N^r\dot{\bar\theta_r}\right)\theta^r(\theta^0)'
    -\frac{N^r}N\dot{\bar\theta_r}\theta^r(\theta^0)'\right]
    =\bar\varepsilon\left[-\dot P_N\theta^0+\dot{\bar\theta_0}\theta^r(\theta^0)'\right];\\
\label{delPNr}
\delta P_{N^r}&=&\bar\varepsilon\left[-\dot P_{N^r}\theta^0+\dot{\bar\theta_r}\theta^r(\theta^0)'\right].
\end{eqnarray}
Here we also used the Hamiltonian equation
$\dot P_N=-\displaystyle\frac{\delta H}{\delta N}
=-\displaystyle\frac{\partial{\cal H}}{\partial N}+\left(\displaystyle\frac{\partial{\cal H}}{\partial N'}\right)'$ (\ref{DPN}). Keeping in mind the relations between $P_N$, $P_{N^r}$ and $\lambda_0$, $\lambda_r$ (\ref{lambda}), one can be convinced that Eqs. (\ref{dell0}), (\ref{dellr}) are correct.

One can find transformations for $\delta P_V$:
\begin{eqnarray}
\delta P_V&=&\bar\varepsilon\left[-\dot P_V\theta^0-P'_V\theta^r
   -P'_N\frac{\partial f}{\partial V}\theta^r
   -P'_{N^r}\frac{\partial f^r}{\partial V}\theta^r
   +P_N\frac{\partial^2 f}{\partial V^2}V(\theta^r)'
   +P_{N^r}\frac{\partial^2 f^r}{\partial V^2}V(\theta^r)'\right.\nonumber\\
&+&P_N\frac{\partial^2 f}{\partial V^2}V N^r(\theta^0)'
   +P_{N^r}\frac{\partial^2 f^r}{\partial V^2}V N^r(\theta^0)'
   +\frac{\partial f^r}{\partial V}\bar P_{\theta^r}(\theta^0)'\theta^r\nonumber\\
\label{delPV}
&+&\left.\frac1N\frac{\partial f}{\partial V}\bar P_{\theta^0}(\theta^0)'\theta^r
   -\frac{N^r}N\frac{\partial f}{\partial V}\bar P_{\theta^r}(\theta^0)'\theta^r
   -\frac{N W W'}{V^2}(\theta^0)'\right].
\end{eqnarray}
The transformation of (\ref{PV}) gives the same result. Similarly one can obtain transformations in extended phase space for $P_W$ and ghosts momenta. They are in correspondence with (\ref{PW}) -- (\ref{Ptr}).

\section{Concluding remarks}
In the present paper we have constructed a self-consistent Hamiltonian dynamics for the generalized spherically symmetric model in extended phase space. Our starting point was the Faddeev -- Popov effective action with gauge-fixing and ghost terms. Thanks to introducing the missing velocities into the Lagrangian by gauge conditions of special form we do not need to invent some prescription how to construct a Hamiltonian function. Hamiltonian equations are proved to be equivalent to the Lagrangian set of equations. The group of transformations in extended phase space includes the group of gauge transformations for all gravitational degrees of freedom. We also have a clear algorithm how to construct a generator of transformations in extended phase space in accordance with the Noether theorem. The necessary condition for the algorithm to work is to find a BRST invariant form of the action. For the present model we have found the additional terms (\ref{add-act}) that guarantees the required BRST invariance. The form of these terms gives us a hint what a BRST invariant form of the action would be in the full gravitational theory. Let us emphasize once again that we do not impose any additional conditions to ensure BRST invariance.

In our opinion, the proposed approach to construct Hamiltonian dynamics for gravity (and, in general, to any constrained system) is of interest by itself, as an alternative to the Dirac approach. On the other hand, it can be considered as a preliminary step to subsequent quantization of the model, and it will be a goal of our further research.

\appendix
\renewcommand{\thesection}{Appendix \Alph{section}.}
\renewcommand{\theequation}{\Alph{section}.\arabic{equation}}

\section{The extended set of Lagrangian equations}
Variation of the effective action (\ref{full-act2}) with respect to $N$, $N^r$ gives the constraints in the Lagrangian formalism which are equivalent to $\left(0\atop 0\right)$ and $\left(0\atop 1\right)$ Einstein equations:
$$
\frac{\partial L}{\partial N}=\partial_0\frac{\partial L}{\partial\dot N}+\partial_r\frac{\partial L}{\partial N'};
$$
\begin{eqnarray}
0&=&\frac{\dot V W\dot W}{N^2}+\frac{V\dot W^2}{2N^2}
   +\frac V2-\frac{(W')^2}{2V}-\frac{W W''}V+\frac{V'W'W}{V^2}\nonumber\\
&-&\frac{W'\dot W V N^r}{N^2}- \frac{W\dot W V(N^r)'}{N^2}
   -\frac{WW'\dot V N^r}{N^2}-\frac{W\dot W V'N^r}{N^2}\nonumber\\
&+&\frac{(W')^2V(N^r)^2}{2N^2}+\frac{W W'V'(N^r)^2}{N^2}+\frac{W W'V N^r(N^r)'}{N^2}\nonumber\\
\label{eLN}
&+&\dot{\pi}_N-\dot{\bar\theta_0}\dot{\theta}^0
   +(\dot{\bar\theta_0})'\theta^r+\dot{\bar\theta_0}(\theta^r)'
   +\dot{\bar\theta_0}N^r(\theta^0)'+\dot{\bar\theta_r}\frac{2N}{V^2}(\theta^0)';
\end{eqnarray}
$$
\frac{\partial L}{\partial N^r}
 =\partial_0\frac{\partial L}{\partial\dot{N^r}}+\partial_r\frac{\partial L}{\partial(N^r)'};
$$
\begin{eqnarray}
0&=&\frac{W W'N'V N^r}{N^2}-\frac{W W'\dot V}N-\frac{W W''V N^r}N+\frac{W V\dot W'}N
   -\frac{N'WV\dot W }{N^2}+\frac{W W'V'N^r}N\nonumber\\
\label{eLNr}
&-&\dot{\pi}_{N^r}+\dot{\bar\theta_r}\dot{\theta^0}-(\dot{\bar\theta_r})'\theta^r
   -2\dot{\bar\theta_r}(\theta^r)'-2\dot{\bar\theta_r}N^r(\theta^0)'
   -\dot{\bar\theta_0}\frac{\partial f}{\partial V}V(\theta^0)'
   -\dot{\bar\theta_r}\frac{\partial f^r}{\partial V}V(\theta^0)'
   -\dot{\bar\theta_0}N(\theta^0)'.
\end{eqnarray}
Variations with respect to $V$, $W$ leads to the equations which are equivalent to dynamical $\left(1\atop 1\right)$ and $\left(2\atop 2\right)$ Einstein equations:
$$
\frac{\partial L}{\partial V}
 =\partial_0\frac{\partial L}{\partial\dot V}+\partial_r\frac{\partial L}{\partial V'};
$$
\begin{eqnarray}
0&=&\frac{\dot W^2}{2N}+\frac{W\ddot W}N-\frac{W\dot W\dot N}{N^2}+\frac N2
   -\frac{N(W')^2}{2V^2}-\frac{N'W'W}{V^2}\nonumber\\
&-&\frac{N'W W'(N^r)^2}{N^2}+\frac{W W''(N^r)^2}N-\frac{W'\dot W N^r}N
   +\frac{WW'N^r\dot N}{N^2}+\frac{N'N^r W\dot W}{N^2}\nonumber\\
&-&\frac{2W\dot W'N^r}N-\frac{WW'\dot N^r}N+\frac{(W')^2(N^r)^2}{2N}+\frac{WW'N^r(N^r)'}N\nonumber\\
&-&\dot\pi_N\frac{\partial f}{\partial V}-\dot\pi_{N^r}\frac{\partial f^r}{\partial V}
   -(\dot{\bar\theta_0})'\frac{\partial f}{\partial V}\theta^r
   +\dot{\bar\theta_0}\frac{\partial^2 f}{\partial V^2}V N^r(\theta^0)'\nonumber\\
&+&\dot{\bar\theta_0}\frac{\partial f}{\partial V}N^r(\theta^0)'
   +\dot{\bar\theta_0}\frac{\partial^2 f}{\partial V^2}V(\theta^r)'
   -\dot{\bar\theta_r}\frac{2N^2}{V^3}(\theta^0)'\nonumber\\
\label{eLV}
&+&\dot{\bar\theta_r}\frac{\partial^2 f^r}{\partial V^2}V N^r(\theta^0)'
   +\dot{\bar\theta_r}\frac{\partial f^r}{\partial V}N^r(\theta^0)'
   +\dot{\bar\theta_r}\frac{\partial^2 f^r}{\partial V^2}V(\theta^r)'
   -(\dot{\bar\theta_r})'\frac{\partial f^r}{\partial V}\theta^r;
\end{eqnarray}
$$
\frac{\partial L}{\partial W}
 =\partial_0\frac{\partial L}{\partial\dot W}+\partial_r\frac{\partial L}{\partial W'};
$$
\begin{eqnarray}
0&=&\frac{W\ddot V}N+\frac{\dot V\dot W}N+\frac{V\ddot W}N-\frac{\dot N\dot V W}{N^2}
   -\frac{\dot N\dot W V}{N^2}-\frac{W N''}V+\frac{W N'V'}{V^2}
   -\frac{W'N'}V+\frac{W'N V'}{V^2}-\frac{W''N}V\nonumber\\
&-&\frac{2V\dot W'N^r}N-\frac{W'\dot V N^r}N-\frac{W'V\dot N^r}N+\frac{W'V N^r\dot N}{N^2}
   +\frac{\dot W V N^r N'}{N^2}-\frac{\dot W V'N^r}N-\frac{2W\dot V'N^r}N\nonumber\\
&-&\frac{WV'\dot N^r}N+\frac{WV'N^r\dot N}{N^2}-\frac{2W\dot V(N^r)'}N+\frac{W\dot V N^r N'}{N^2}
   -\frac{\dot W V(N^r)'}N-\frac{W V(\dot N^r)'}N+\frac{WV\dot N(N^r)'}{N^2}\nonumber\\
&+&\frac{W V''(N^r)^2}N+\frac{3W V'N^r(N^r)'}N-\frac{W V'(N^r)^2 N'}{N^2}
   +\frac{W V((N^r)')^2}N+\frac{W V N^r(N^r)''}N\nonumber\\
&-&\frac{W V N'N^r(N^r)'}{N^2}+\frac{W''V(N^r)^2}N
   +\frac{W'V'(N^r)^2}N+\frac{2W'V N^r(N^r)'}N-\frac{W'V(N^r)^2 N'}{N^2}\nonumber\\
&-&\dot\pi_N\frac{\partial f}{\partial W}-\dot\pi_{N^r}\frac{\partial f^r}{\partial W}
   -(\dot{\bar\theta_0})'\frac{\partial f}{\partial W}\theta^r
   -\dot{\bar\theta_0}\frac{\partial f}{\partial W}(\theta^r)'
   +\dot{\bar\theta_0}\frac{\partial^2 f}{\partial V\partial W}V N^r(\theta^0)'
   +\dot{\bar\theta_0}\frac{\partial^2 f}{\partial V\partial W}V(\theta^r)'\nonumber\\
\label{eLW}
&+&\dot{\bar\theta_r}\frac{\partial^2 f^r}{\partial V\partial W}V N^r(\theta^0)'
   +\dot{\bar\theta_r}\frac{\partial^2 f^r}{\partial V\partial W}V(\theta^r)'
   -(\dot{\bar\theta_r})'\frac{\partial f^r}{\partial W}\theta^r
   -\dot{\bar\theta_r}\frac{\partial f^r}{\partial W}(\theta^r)'.
\end{eqnarray}

We also have four equations for two pairs of ghosts:
$$
\frac{\partial L}{\partial\bar\theta_0}
 =\partial_0\frac{\partial L}{\partial\dot{\bar\theta_0}}
 +\partial_r\frac{\partial L}{\partial(\bar\theta_0)'};
$$
\begin{eqnarray}
0&=&N'\dot\theta^r-\frac{\partial f}{\partial V}V'\dot\theta^r
   -\frac{\partial f}{\partial W}W'\dot\theta^r+\dot N'\theta^r
   -\frac{\partial^2 f}{\partial V^2}\dot V V'\theta^r\nonumber\\
&-&\frac{\partial^2 f}{\partial V\partial W}V'\dot W\theta^r
   -\frac{\partial f}{\partial V}\dot V'\theta^r
   -\frac{\partial^2 f}{\partial V\partial W}\dot V W'\theta^r
   -\frac{\partial^2 f}{\partial W^2}\dot W W'\theta^r
   -\frac{\partial f}{\partial W}\dot W'\theta^r\nonumber\\
&+&\dot N\dot\theta^0+N\ddot\theta^0-\dot N N^r(\theta^0)'-N\dot N^r(\theta^0)'-NN^r(\dot\theta^0)'
   -\frac{\partial^2 f}{\partial V^2}\dot V V N^r(\theta^0)'\nonumber\\
&-&\frac{\partial^2 f}{\partial V\partial W}\dot W V N^r(\theta^0)'
   -\frac{\partial f}{\partial V}\dot V N^r(\theta^0)'
   -\frac{\partial f}{\partial V}V\dot N^r(\theta^0)'
   -\frac{\partial f}{\partial V}V N^r(\dot\theta^0)'\nonumber\\
\label{eLbt0}
&-&\frac{\partial^2 f}{\partial V^2}\dot V V(\theta^r)'
   -\frac{\partial^2 f}{\partial V\partial W}\dot W V(\theta^r)'
   -\frac{\partial f}{\partial V}\dot V(\theta^r)'
   -\frac{\partial f}{\partial V}V(\dot\theta^r)';
\end{eqnarray}
$$
\frac{\partial L}{\partial\theta^0}
 =\partial_0\frac{\partial L}{\partial\dot\theta^0}+\partial_r\frac{\partial L}{\partial(\theta^0)'};
$$
\begin{eqnarray}
0&=&\ddot{\bar\theta_0}N+\dot{\bar\theta_0}\dot N
   -(\dot{\bar\theta_0})'N N^r
   -\dot{\bar\theta_0}N'N^r-\dot{\bar\theta_0}N(N^r)'\nonumber\\
&-&(\dot{\bar\theta_0})'\frac{\partial f}{\partial V}V N^r
   -\dot{\bar\theta_0}\frac{\partial^2 f}{\partial V^2}V'V N^r
   -\dot{\bar\theta_0}\frac{\partial^2 f}{\partial V\partial W}W'V N^r
   -\dot{\bar\theta_0}\frac{\partial f}{\partial V}V'N^r
   -\dot{\bar\theta_0}\frac{\partial f}{\partial V}V(N^r)'\nonumber\\
&+&\ddot{\bar\theta_r}N^r+\dot{\bar\theta_r}\dot N^r
   -(\dot{\bar\theta_r})'\left(\frac{N^2}{V^2}+(N^r)^2\right)
   -2\dot{\bar\theta_r}\;\frac{N N'}{V^2}
   +2\dot{\bar\theta_r}\;\frac{N^2V'}{V^3}
   -2\dot{\bar\theta_r}N^r(N^r)'\nonumber\\
\label{eLt0}
&-&(\dot{\bar\theta_r})'\frac{\partial f^r}{\partial V}V N^r
   -\dot{\bar\theta_r}\frac{\partial^2 f^r}{\partial V^2}V'V N^r
   -\dot{\bar\theta_r}\frac{\partial^2 f^r}{\partial V\partial W}W'V N^r
   -\dot{\bar\theta_r}\frac{\partial f^r}{\partial V}V'N^r
   -\dot{\bar\theta_r}\frac{\partial f^r}{\partial V}V(N^r)';
\end{eqnarray}
$$
\frac{\partial L}{\partial\bar\theta_r}
 =\partial_0\frac{\partial L}{\partial\dot{\bar\theta_r}}
 +\partial_r\frac{\partial L}{\partial(\bar\theta_r)'};
$$
\begin{eqnarray}
0&=&\dot N^r\dot\theta^0+N^r\ddot\theta^0-2\frac{N\dot N}{V^2}(\theta^0)'
   +2\frac{N^2\dot V}{V^3}(\theta^0)'-\frac{N^2}{V^2}(\dot\theta^0)'
   -2N^r\dot N^r(\theta^0)'\nonumber\\
&-&(N^r)^2(\dot\theta^0)'+\ddot\theta^r-\dot N^r(\theta^r)'-N^r(\dot\theta^r)'
   +(\dot N^r)'\theta^r+(N^r)'\dot\theta^r\nonumber\\
&-&\frac{\partial^2 f^r}{\partial V^2}\dot V V N^r(\theta^0)'
   -\frac{\partial^2 f^r}{\partial V^2}\dot V V(\theta^r)'
   -\frac{\partial^2 f^r}{\partial V^2}\dot V V'\theta^r\nonumber\\
&-&\frac{\partial^2 f^r}{\partial V\partial W}\dot W V N^r(\theta^0)'
   -\frac{\partial^2 f^r}{\partial V\partial W}\dot W V(\theta^r)'
   -\frac{\partial^2 f^r}{\partial V\partial W}\dot W V'\theta^r\nonumber\\
&-&\frac{\partial f^r}{\partial V}\dot V N^r(\theta^0)'
   -\frac{\partial f^r}{\partial V}V\dot N^r(\theta^0)'
   -\frac{\partial f^r}{\partial V}V N^r(\dot\theta^0)'
   -\frac{\partial f^r}{\partial V}\dot V(\theta^r)'
   -\frac{\partial f^r}{\partial V}V(\dot\theta^r)'\nonumber\\
\label{eLbtr}
&-&\frac{\partial f^r}{\partial V}\dot V'\theta^r
   -\frac{\partial f^r}{\partial V}V'\dot\theta^r
   -\frac{\partial^2 f^r}{\partial V\partial W}\dot V W'\theta^r
   -\frac{\partial^2 f^r}{\partial W^2}\dot W W'\theta^r
   -\frac{\partial f^r}{\partial W}\dot W'\theta^r
   -\frac{\partial f^r}{\partial W}W'\dot\theta^r;
\end{eqnarray}
$$
\frac{\partial L}{\partial\theta^r}
 =\partial_0\frac{\partial L}{\partial\dot\theta^r}+\partial_r\frac{\partial L}{\partial(\theta^r)'};
$$
\begin{eqnarray}
0&=&\dot{\bar\theta_0}N'-\dot{\bar\theta_0}\frac{\partial f}{\partial W}W'
   +(\dot{\bar\theta_0})'\frac{\partial f}{\partial V}V
   +\dot{\bar\theta_0}\frac{\partial^2 f}{\partial V^2}V'V
   +\dot{\bar\theta_0}\frac{\partial^2 f}{\partial V\partial W}W'V\nonumber\\
&-&\ddot{\bar\theta_r}+(\dot{\bar\theta_r})'N^r+2\dot{\bar\theta_r}(N^r)'
   +(\dot{\bar\theta_r})'\frac{\partial f^r}{\partial V}V\nonumber\\
\label{eLtr}
&+&\dot{\bar\theta_r}\frac{\partial^2 f^r}{\partial V^2}V'V
   +\dot{\bar\theta_r}\frac{\partial^2 f^r}{\partial V\partial W}W'V
   -\dot{\bar\theta_r}\frac{\partial f^r}{\partial W}W'.
\end{eqnarray}
Variation with respect to $\pi_N$, $\pi_{N^r}$ yields the gauge conditions (\ref{d.c.1}), (\ref{d.c.2}):
\begin{eqnarray}
\label{d.c.11}
\dot N&=&\frac{\partial f}{\partial V}\dot V+\frac{\partial f}{\partial W}\dot W;\\
\label{d.c.22}
\dot N^r&=&\frac{\partial f^r}{\partial V}\dot V+\frac{\partial f^r}{\partial W}\dot W.
\end{eqnarray}
The equations (\ref{eLN}) -- (\ref{d.c.22}) form the extended set of Lagrangian equations for the generalized spherically symmetric gravitational model.

\section{The set of Hamiltonian equations in extended phase space}
In this Appendix we present the full set of Hamiltonian equations in extended phase space.
\begin{eqnarray}
\dot N&=&\frac{\partial f}{\partial V}V'N^r
   +\frac{\partial f}{\partial W}W'N^r
   +\frac{\partial f}{\partial V}V(N^r)'
   +\frac N W P_V\frac{\partial f}{\partial W}
   +\frac N W P_W\frac{\partial f}{\partial V}
   -\frac{N V}{W^2}P_V\frac{\partial f}{\partial V}\nonumber\\
\label{DN}
&-&\frac{N V}{W^2}P_N\left(\frac{\partial f}{\partial V}\right)^2
   +\frac{2N}W P_N\frac{\partial f}{\partial V}\frac{\partial f}{\partial W}
   +\frac N W P_{N^r}\frac{\partial f}{\partial V}\frac{\partial f^r}{\partial W}
   +\frac N W P_{N^r}\frac{\partial f}{\partial W}\frac{\partial f^r}{\partial V}
   -\frac{N V}{W^2}P_{N^r}\frac{\partial f}{\partial V}\frac{\partial f^r}{\partial V};\\
\dot P_N&=&-\frac1W P_V P_W+\frac V{2W^2}P_V^2+\frac{W''W}V-\frac{V'W'W}{V^2}
   +\frac{(W')^2}{2V}-\frac V2\nonumber\\
&-&\frac1W P_V P_N\frac{\partial f}{\partial W}
   -\frac1W P_W P_N\frac{\partial f}{\partial V}
   +\frac V{W^2}P_V P_N\frac{\partial f}{\partial V}
   -\frac1W P_V P_{N^r}\frac{\partial f^r}{\partial W}
   -\frac1W P_W P_{N^r}\frac{\partial f^r}{\partial V}\nonumber\\
&+&\frac V{W^2}P_V P_{N^r}\frac{\partial f^r}{\partial V}
   +\frac V{2W^2} P_N^2\left(\frac{\partial f}{\partial V}\right)^2
   +\frac V{2W^2}P_{N^r}^2\left(\frac{\partial f^r}{\partial V}\right)^2
   -\frac1W P_N^2\frac{\partial f}{\partial V}\frac{\partial f}{\partial W}
   -\frac1W P_{N^r}^2\frac{\partial f^r}{\partial V}\frac{\partial f^r}{\partial W}\nonumber\\
&-&\frac1W P_N P_{N^r}\frac{\partial f}{\partial V}\frac{\partial f^r}{\partial W}
   -\frac1W P_N P_{N^r}\frac{\partial f}{\partial W}\frac{\partial f^r}{\partial V}
   +\frac V{W^2}P_N P_{N^r}\frac{\partial f}{\partial V}\frac{\partial f^r}{\partial V}\nonumber\\
&+&\frac1{N^2}\bar P_{\theta^0}P_{\bar\theta_0}
   -\frac{N^r}{N^2}\bar P_{\theta^r}P_{\bar\theta_0}
   -\frac1N(\bar P_{\theta^0})'\theta^r-\frac1N\bar P_{\theta^0}(\theta^r)'\nonumber\\
&+&\frac{(N^r)'}N\bar P_{\theta^r}\theta^r
   +\frac{N^r}N(\bar P_{\theta^r})'\theta^r
   +\frac{N^r}N\bar P_{\theta^r}(\theta^r)'
   -\frac{2N}{V^2}\bar P_{\theta^r}(\theta^0)'\nonumber\\
&+&\frac V{N^2}\frac{\partial f}{\partial V}\bar P_{\theta^0}(\theta^r)'
   +\frac{V N^r}{N^2}\frac{\partial f}{\partial V}\bar P_{\theta^0}(\theta^0)'
   -\frac{V N^r}{N^2}\frac{\partial f}{\partial V}\bar P_{\theta^r}(\theta^r)'
   -\frac{V(N^r)^2}{N^2}\frac{\partial f}{\partial V}\bar P_{\theta^r}(\theta^0)'\nonumber\\
\label{DPN}
&+&\frac{V'}{N^2}\frac{\partial f}{\partial V}\bar P_{\theta^0}\theta^r
   -\frac{V'N^r}{N^2}\frac{\partial f}{\partial V}\bar P_{\theta^r}\theta^r
   +\frac{W'}{N^2}\frac{\partial f}{\partial W}\bar P_{\theta^0}\theta^r
   -\frac{W'N^r}{N^2}\frac{\partial f}{\partial W}\bar P_{\theta^r}\theta^r;\\
\dot N^r&=&\frac{\partial f^r}{\partial V}V'N^r
   +\frac{\partial f^r}{\partial W}W'N^r
   +\frac{\partial f^r}{\partial V}V(N^r)'
   +\frac N W P_V\frac{\partial f^r}{\partial W}
   +\frac N W P_W\frac{\partial f^r}{\partial V}
   -\frac{N V}{W^2}P_V\frac{\partial f^r}{\partial V}\nonumber\\
&-&\frac{N V}{W^2}P_{N^r}\left(\frac{\partial f^r}{\partial V}\right)^2
   +\frac{2N}W P_{N^r}\frac{\partial f^r}{\partial V}\frac{\partial f^r}{\partial W}
   +\frac N W P_N\frac{\partial f}{\partial V}\frac{\partial f^r}{\partial W}\nonumber\\
\label{DNr}
&+&\frac N W P_N\frac{\partial f}{\partial W}\frac{\partial f^r}{\partial V}
   -\frac{N V}{W^2}P_N\frac{\partial f}{\partial V}\frac{\partial f^r}{\partial V}';\\
\dot P_{N^r}&=&(P_V)'V-P_W W'
   -P_N\frac{\partial f}{\partial W}W'
   +(P_N)'\frac{\partial f}{\partial V}V
   +P_N\frac{\partial^2 f}{\partial V^2}V V'
   +P_N\frac{\partial^2 f}{\partial V\partial W}V W'\nonumber\\
&-&P_{N^r}\frac{\partial f^r}{\partial W}W'
   +(P_{N^r})'\frac{\partial f^r}{\partial V}V
   +P_{N^r}\frac{\partial^2 f^r}{\partial V^2}V V'
   +P_{N^r}\frac{\partial^2 f^r}{\partial V\partial W}V W'\nonumber\\
&+&\frac1N\bar P_{\theta^r}P_{\bar\theta_0}
   -\bar P_{\theta^0}(\theta^0)'
   -2\bar P_{\theta^r}(\theta^r)'
   -(\bar P_{\theta^r})'\theta^r
   -\frac{N'}N\bar P_{\theta^r}\theta^r\nonumber\\
&-&\frac{\partial f^r}{\partial V}V\bar P_{\theta^r}(\theta^0)'
   -\frac V N\frac{\partial f}{\partial V}\bar P_{\theta^0}(\theta^0)'
   +\frac V N\frac{\partial f}{\partial V}\bar P_{\theta^r}(\theta^r)'\nonumber\\
\label{DPNr}
&+&\frac{2V N^r}N\frac{\partial f}{\partial V}\bar P_{\theta^r}(\theta^0)'
   +\frac{V'}N\frac{\partial f}{\partial V}\bar P_{\theta^r}\theta^r
   +\frac{W'}N\frac{\partial f}{\partial W}\bar P_{\theta^r}\theta^r.
\end{eqnarray}
It can be checked by direct calculations that Eqs. (\ref{DPN}), (\ref{DPNr}) coincide with the constraints equations in the Lagrangian formalism (\ref{eLN}), (\ref{eLNr}), while Eqs. (\ref{DN}), (\ref{DNr}) correspond to the gauge conditions (\ref{d.c.11}), (\ref{d.c.22}). The other Hamiltonian equations are:
\begin{eqnarray}
\dot V&=&\frac N W P_W-\frac{N V}{W^2}P_V+V'N^r+V(N^r)'\nonumber\\
\label{DV}
&+&\frac N W P_N\frac{\partial f}{\partial W}
   -\frac{N V}{W^2}P_N\frac{\partial f}{\partial V}
   +\frac N W P_{N^r}\frac{\partial f^r}{\partial W}
   -\frac{N V}{W^2}P_{N^r}\frac{\partial f^r}{\partial V};\\
\dot P_V&=&\frac N{2W^2}P_V^2+(P_V)'N^r+\frac{N'W'W}{V^2}
   +\frac{N(W')^2}{2V^2}-\frac N2\nonumber\\
&+&(P_N)'\frac{\partial f}{\partial V}N^r
   -P_N\frac{\partial^2 f}{\partial V^2}V(N^r)'
   +(P_{N^r})'\frac{\partial f^r}{\partial V}N^r
   -P_{N^r}\frac{\partial^2 f^r}{\partial V^2}V(N^r)'\nonumber\\
&-&\frac N W P_V P_N\frac{\partial^2 f}{\partial V\partial W}
   -\frac N W P_W P_N\frac{\partial^2 f}{\partial V^2}
   +\frac N{W^2}P_V P_N\frac{\partial f}{\partial V}
   +\frac{N V}{W^2}P_V P_N\frac{\partial^2 f}{\partial V^2}\nonumber\\
&-&\frac N W P_V P_{N^r}\frac{\partial^2 f^r}{\partial V\partial W}
   -\frac N W P_W P_{N^r}\frac{\partial^2 f^r}{\partial V^2}
   +\frac N{W^2}P_V P_{N^r}\frac{\partial f^r}{\partial V}
   +\frac{N V}{W^2}P_V P_{N^r}\frac{\partial^2 f^r}{\partial V^2}\nonumber\\
&+&\frac N{2W^2} P_N^2\left(\frac{\partial f}{\partial V}\right)^2
   +\frac{N V}{W^2} P_N^2\frac{\partial f}{\partial V}\frac{\partial^2 f}{\partial V^2}
   +\frac N{2W^2}P_{N^r}^2\left(\frac{\partial f^r}{\partial V}\right)^2
   +\frac{N V}{W^2}P_{N^r}^2\frac{\partial f^r}{\partial V}
    \frac{\partial^2 f^r}{\partial V^2}\nonumber\\
&-&\frac N W P_N^2\frac{\partial^2 f}{\partial V^2}
    \frac{\partial f}{\partial W}
   -\frac N W P_N^2\frac{\partial f}{\partial V}
    \frac{\partial^2 f}{\partial V\partial W}
   -\frac N W P_{N^r}^2\frac{\partial^2 f^r}{\partial V^2}
    \frac{\partial f^r}{\partial W}
   -\frac N W P_{N^r}^2\frac{\partial f^r}{\partial V}
    \frac{\partial^2 f^r}{\partial V\partial W}\nonumber\\
&-&\frac N W P_N P_{N^r}\frac{\partial^2 f}{\partial V^2}
    \frac{\partial f^r}{\partial W}
   -\frac N W P_N P_{N^r}\frac{\partial f}{\partial V}
    \frac{\partial^2 f^r}{\partial V\partial W}
   -\frac N W P_N P_{N^r}\frac{\partial^2 f}{\partial V\partial W}
    \frac{\partial f^r}{\partial V}
   -\frac N W P_N P_{N^r}\frac{\partial f}{\partial W}
    \frac{\partial^2 f^r}{\partial V^2}\nonumber\\
&+&\frac N{W^2}P_N P_{N^r}\frac{\partial f}{\partial V}
    \frac{\partial f^r}{\partial V}
   +\frac{N V}{W^2}P_N P_{N^r}\frac{\partial^2 f}{\partial V^2}
    \frac{\partial f^r}{\partial V}
   +\frac{N V}{W^2}P_N P_{N^r}\frac{\partial f}{\partial V}
    \frac{\partial^2 f^r}{\partial V^2}\nonumber\\
&+&\frac{2N^2}{V^3}\bar P_{\theta^r}(\theta^0)'
   +\frac{\partial f^r}{\partial V}(\bar P_{\theta^r})'\theta^r
   -\frac{\partial^2 f^r}{\partial V^2}V N^r\bar P_{\theta^r}(\theta^0)'
   -\frac{\partial f^r}{\partial V}N^r\bar P_{\theta^r}(\theta^0)'
   -\frac{\partial^2 f^r}{\partial V^2}V\bar P_{\theta^r}(\theta^r)'\nonumber\\
&-&\frac V N\frac{\partial^2 f}{\partial V^2}\bar P_{\theta^0}(\theta^r)'
   -\frac{N^r}N\frac{\partial f}{\partial V}\bar P_{\theta^0}(\theta^0)'
   -\frac{V N^r}N\frac{\partial^2 f}{\partial V^2}\bar P_{\theta^0}(\theta^0)'
   +\frac{V N^r}N\frac{\partial^2 f}{\partial V^2}\bar P_{\theta^r}(\theta^r)'\nonumber\\
&+&\frac{(N^r)^2}N\frac{\partial f}{\partial V}\bar P_{\theta^r}(\theta^0)'
   +\frac{V(N^r)^2}N\frac{\partial^2 f}{\partial V^2}\bar P_{\theta^r}(\theta^0)'
   -\frac{N'}{N^2}\frac{\partial f}{\partial V}\bar P_{\theta^0}\theta^r
   +\frac1N\frac{\partial f}{\partial V}(\bar P_{\theta^0})'\theta^r\nonumber\\
\label{DPV}
&-&\frac{(N^r)'}N\frac{\partial f}{\partial V}\bar P_{\theta^r}\theta^r
   +\frac{N'N^r}{N^2}\frac{\partial f}{\partial V}\bar P_{\theta^r}\theta^r
   -\frac{N^r}N\frac{\partial f}{\partial V}(\bar P_{\theta^r})'\theta^r;\\
\label{DW}
\dot W&=&\frac N W P_V+W'N^r
   +\frac N W P_N\frac{\partial f}{\partial V}
   +\frac N W P_{N^r}\frac{\partial f^r}{\partial V};\\
\dot P_W&=&\frac N{W^2}P_V P_W-\frac{N V}{W^3}P_V^2+(P_W)'N^r+P_W(N^r)'
   +\frac{N''W}V-\frac{N'V'W}{V^2}+\frac{N'W'}V+\frac{N W''}V-\frac{N V'W'}{V^2}\nonumber\\
&+&(P_N)'\frac{\partial f}{\partial W}N^r
   +P_N\frac{\partial f}{\partial W}(N^r)'
   -P_N\frac{\partial^2 f}{\partial V\partial W}V(N^r)'
   +(P_{N^r})'\frac{\partial f^r}{\partial W}N^r
   +P_{N^r}\frac{\partial f^r}{\partial W}(N^r)'\nonumber\\
&-&P_{N^r}\frac{\partial^2 f^r}{\partial V\partial W}V(N^r)'
   +\frac N{W^2}P_V P_N\frac{\partial f}{\partial W}
   -\frac N W P_V P_N\frac{\partial^2 f}{\partial W^2}
   +\frac N{W^2}P_W P_N\frac{\partial f}{\partial V}
   -\frac N W P_W P_N\frac{\partial^2 f}{\partial V\partial W}\nonumber\\
&-&\frac{2N V}{W^3}P_V P_N\frac{\partial f}{\partial V}
   +\frac{N V}{W^2}P_V P_N\frac{\partial^2 f}{\partial V\partial W}
   +\frac N{W^2}P_V P_{N^r}\frac{\partial f^r}{\partial W}
   -\frac N W P_V P_{N^r}\frac{\partial^2 f^r}{\partial W^2}\nonumber\\
&+&\frac N{W^2}P_W P_{N^r}\frac{\partial f^r}{\partial V}
   -\frac N W P_W P_{N^r}\frac{\partial^2 f^r}{\partial V\partial W}
   -\frac{2N V}{W^3}P_V P_{N^r}\frac{\partial f^r}{\partial V}
   +\frac{N V}{W^2}P_V P_{N^r}\frac{\partial^2 f^r}{\partial V\partial W}\nonumber\\
&-&\frac{N V}{W^3}P_N^2\left(\frac{\partial f}{\partial V}\right)^2
   +\frac{N V}{W^2}P_N^2\frac{\partial f}{\partial V}
    \frac{\partial^2 f}{\partial V\partial W}
   -\frac{N V}{W^3}P_{N^r}^2\left(\frac{\partial f^r}{\partial V}\right)^2
   +\frac{N V}{W^2}P_{N^r}^2\frac{\partial f^r}{\partial V}
    \frac{\partial^2 f^r}{\partial V\partial W}\nonumber\\
&+&\frac N{W^2}P_N^2\frac{\partial f}{\partial V}
    \frac{\partial f}{\partial W}
   -\frac N W P_N^2\frac{\partial^2 f}{\partial V\partial W}
    \frac{\partial f}{\partial W}
   -\frac N W P_N^2\frac{\partial f}{\partial V}
    \frac{\partial^2 f}{\partial W^2}\nonumber\\
&+&\frac N{W^2}P_{N^r}^2\frac{\partial f^r}{\partial V}
    \frac{\partial f^r}{\partial W}
   -\frac N W P_{N^r}^2\frac{\partial^2 f^r}{\partial V\partial W}
    \frac{\partial f^r}{\partial W}
   -\frac N W P_{N^r}^2\frac{\partial f^r}{\partial V}
    \frac{\partial^2 f^r}{\partial W^2}\nonumber\\
&+&\frac N{W^2}P_N P_{N^r}\frac{\partial f}{\partial V}
    \frac{\partial f^r}{\partial W}
   -\frac N W P_N P_{N^r}\frac{\partial^2 f}{\partial V\partial W}
    \frac{\partial f^r}{\partial W}
   -\frac N W P_N P_{N^r}\frac{\partial f}{\partial V}
    \frac{\partial^2 f^r}{\partial W^2}\nonumber\\
&+&\frac N{W^2}P_N P_{N^r}\frac{\partial f}{\partial W}
    \frac{\partial f^r}{\partial V}
   -\frac N W P_N P_{N^r}\frac{\partial^2 f}{\partial W^2}
    \frac{\partial f^r}{\partial V}
   -\frac N W P_N P_{N^r}\frac{\partial f}{\partial W}
    \frac{\partial^2 f^r}{\partial V\partial W}\nonumber\\
&-&\frac{2N V}{W^3}P_N P_{N^r}\frac{\partial f}{\partial V}
    \frac{\partial f^r}{\partial V}
   +\frac{N V}{W^2}P_N P_{N^r}\frac{\partial^2 f}{\partial V\partial W}
    \frac{\partial f^r}{\partial V}
   +\frac{N V}{W^2}P_N P_{N^r}\frac{\partial f}{\partial V}
    \frac{\partial^2 f^r}{\partial V\partial W}\nonumber\\
&+&\frac{\partial f^r}{\partial W}(\bar P_{\theta^r})'\theta^r
   +\frac{\partial f^r}{\partial W}\bar P_{\theta^r}(\theta^r)'
   -\frac{\partial^2 f^r}{\partial V\partial W}V N^r\bar P_{\theta^r}(\theta^0)'
   -\frac{\partial^2 f^r}{\partial V\partial W}
    V\bar P_{\theta^r}(\theta^r)'\nonumber\\
&-&\frac V N\frac{\partial^2 f}{\partial V\partial W}\bar P_{\theta^0}(\theta^r)'
   -\frac{V N^r}N\frac{\partial^2 f}{\partial V\partial W}\bar P_{\theta^0}(\theta^0)'
   +\frac{V N^r}N\frac{\partial^2 f}{\partial V\partial W}
    \bar P_{\theta^r}(\theta^r)'\nonumber\\
&+&\frac{V(N^r)^2}N\frac{\partial^2 f}{\partial V\partial W}\bar P_{\theta^r}(\theta^0)'
   -\frac{N'}{N^2}\frac{\partial f}{\partial W}\bar P_{\theta^0}\theta^r
   +\frac1N\frac{\partial f}{\partial W}(\bar P_{\theta^0})'\theta^r
   +\frac1N\frac{\partial f}{\partial W}\bar P_{\theta^0}(\theta^r)'\nonumber\\
\label{DPW}
&-&\frac{(N^r)'}N\frac{\partial f}{\partial W}\bar P_{\theta^r}\theta^r
   +\frac{N^r N'}{N^2}\frac{\partial f}{\partial W}\bar P_{\theta^r}\theta^r
   -\frac{N^r}N\frac{\partial f}{\partial W}(\bar P_{\theta^r})'\theta^r
   -\frac{N^r}N\frac{\partial f}{\partial W}\bar P_{\theta^r}(\theta^r)'.
\end{eqnarray}
Eqs. (\ref{DV}) -- (\ref{DPW}) are equivalent to the dynamical Lagrangian equations (\ref{eLV}), (\ref{eLW}).
\begin{eqnarray}
\label{Dt0}
\dot\theta^0&=&\frac1N P_{\bar\theta_0}-\frac{N'}N\theta^r+N^r(\theta^0)'
   +\frac V N\frac{\partial f}{\partial V}(\theta^r)'
   +\frac{V N^r}N\frac{\partial f}{\partial V}(\theta^0)'
   +\frac{V'}N\frac{\partial f}{\partial V}\theta^r
   +\frac{W'}N\frac{\partial f}{\partial W}\theta^r;\\
\label{DPbt0}
\dot P_{\bar\theta_0}&=&0;\\
\label{Dbt0}
\dot{\bar\theta_0}&=&\frac1N\bar P_{\theta^0}-\frac{N^r}N\bar P_{\theta^r};\\
\dot{\bar P}_{\theta^0}&=&(N^r)'\bar P_{\theta^0}+N^r(\bar P_{\theta^0})'
   +\frac{2N N'}{V^2}\bar P_{\theta^r}-\frac{2N^2V'}{V^3}\bar P_{\theta^r}
   +\frac{N^2}{V^2}(\bar P_{\theta^r})'\nonumber\\
&+&\frac{\partial^2 f^r}{\partial V^2}V V'N^r\bar P_{\theta^r}
   +\frac{\partial^2 f^r}{\partial V\partial W}V W'N^r\bar P_{\theta^r}
   +\frac{\partial f^r}{\partial V}V'N^r\bar P_{\theta^r}
   +\frac{\partial f^r}{\partial V}V(N^r)'\bar P_{\theta^r}
   +\frac{\partial f^r}{\partial V}V N^r(\bar P_{\theta^r})'\nonumber\\
&+&\frac{V'N^r}N\frac{\partial f}{\partial V}\bar P_{\theta^0}
   +\frac{V(N^r)'}N\frac{\partial f}{\partial V}\bar P_{\theta^0}
   -\frac{V N'N^r}{N^2}\frac{\partial f}{\partial V}\bar P_{\theta^0}
   +\frac{V V'N^r}N\frac{\partial^2 f}{\partial V^2}\bar P_{\theta^0}
   +\frac{V W'N^r}N\frac{\partial^2 f}{\partial V\partial W}\bar P_{\theta^0}\nonumber\\
&+&\frac{V N^r}N\frac{\partial f}{\partial V}(\bar P_{\theta^0})'
   -\frac{V'(N^r)^2}N\frac{\partial f}{\partial V}\bar P_{\theta^r}
   -\frac{2V N^r(N^r)'}N\frac{\partial f}{\partial V}\bar P_{\theta^r}
   +\frac{V N'(N^r)^2}{N^2}\frac{\partial f}{\partial V}\bar P_{\theta^r}\nonumber\\
\label{DbPt0}
&-&\frac{V V'(N^r)^2}N\frac{\partial^2 f}{\partial V^2}\bar P_{\theta^r}
   -\frac{V W'(N^r)^2}N\frac{\partial^2 f}{\partial V\partial W}\bar P_{\theta^r}
   -\frac{V(N^r)^2}N\frac{\partial f}{\partial V}(\bar P_{\theta^r})';\\
\dot\theta^r&=&P_{\bar\theta_r}-\frac{N^r}NP_{\bar\theta_0}+N^r(\theta^r)'
   -(N^r)'\theta^r+\frac{N'N^r}N\theta^r+\frac{N^2}{V^2}(\theta^0)'\nonumber\\
&+&\frac{\partial f^r}{\partial V}V'\theta^r
   +\frac{\partial f^r}{\partial W}W'\theta^r
   +\frac{\partial f^r}{\partial V}V N^r(\theta^0)'
   +\frac{\partial f^r}{\partial V}V(\theta^r)'\nonumber\\
\label{Dtr}
&-&\frac{V N^r}N\frac{\partial f}{\partial V}(\theta^r)'
   -\frac{V(N^r)^2}N\frac{\partial f}{\partial V}(\theta^0)'
   -\frac{V'N^r}N\frac{\partial f}{\partial V}\theta^r
   -\frac{W'N^r}N\frac{\partial f}{\partial W}\theta^r;\\
\label{DPbtr}
\dot P_{\bar\theta_r}&=&0;\\
\label{Dbtr}
\dot{\bar\theta_r}&=&\bar P_{\theta^r};\\
\dot{\bar P}_{\theta^r}&=&\frac{N'}N\bar P_{\theta^0}
   +2(N^r)'\bar P_{\theta^r}+N^r(\bar P_{\theta^r})'
   -\frac{N'N^r}N\bar P_{\theta^r}
   -\frac{\partial f^r}{\partial W}W'\bar P_{\theta^r}\nonumber\\
&+&\frac{\partial^2 f^r}{\partial V^2}V V'\bar P_{\theta^r}
   +\frac{\partial^2 f^r}{\partial V\partial W}V W'\bar P_{\theta^r}
   +\frac{\partial f^r}{\partial V}V(\bar P_{\theta^r})'
   -\frac{N'V}{N^2}\frac{\partial f}{\partial V}\bar P_{\theta^0}\nonumber\\
&+&\frac{V V'}N\frac{\partial^2 f}{\partial V^2}\bar P_{\theta^0}
   +\frac{V W'}N\frac{\partial^2 f}{\partial V\partial W}\bar P_{\theta^0}
   +\frac V N\frac{\partial f}{\partial V}(\bar P_{\theta^0})'
   -\frac{V(N^r)'}N\frac{\partial f}{\partial V}\bar P_{\theta^r}\nonumber\\
&+&\frac{V N'N^r}{N^2}\frac{\partial f}{\partial V}\bar P_{\theta^r}
   -\frac{V V'N^r}N\frac{\partial^2 f}{\partial V^2}\bar P_{\theta^r}
   -\frac{V W'N^r}N\frac{\partial^2 f}{\partial V\partial W}\bar P_{\theta^r}
   -\frac{V N^r}N\frac{\partial f}{\partial V}(\bar P_{\theta^r})'\nonumber\\
\label{DbPtr}
&-&\frac{W'}N\frac{\partial f}{\partial W}\bar P_{\theta^0}
   +\frac{W'N^r}N\frac{\partial f}{\partial W}\bar P_{\theta^r}.
\end{eqnarray}
Eqs. (\ref{Dt0}) -- (\ref{DbPtr}) are equivalent to the equations for ghosts (\ref{eLbt0}) -- (\ref{eLtr}).

\end{document}